\documentclass[12pt,a4paper,final]{article}
\usepackage[utf8]{inputenc}
\usepackage[english]{babel}
\usepackage{amsmath}
\usepackage{amsfonts}
\usepackage{amssymb}
\usepackage{graphicx}

\usepackage{filecontents}
\usepackage[colorlinks]{hyperref}
\usepackage{amssymb}
\usepackage[table,xcdraw,dvipsnames]{xcolor}
\usepackage{color}

\AtBeginDocument{%
 \hypersetup{
  linkcolor  = NavyBlue,
  citecolor  = OliveGreen,
  urlcolor   = blue,
  colorlinks = true,
}}

\date{}
\author{Jithender J. Timothy\footnote{Research Associate, Institute for structural mechanics, Ruhr University Bochum, 44801, Germany. Ph: ++0049 234 32 29070, email: timothy.jithenderjaswant@rub.de}}
\begin{document}
\title{How does propaganda influence the opinion dynamics of a population ?}

\maketitle
\begin{abstract}
The evolution of opinions in a population of individuals who constantly interact with a common source of user-generated content (i.e. the internet) and are also subject to propaganda is analyzed using computer simulations. The model is based on the bounded confidence approach. In the absence of  propaganda, computer simulations show that the online population as a whole is either fragmented, polarized or in perfect harmony on a certain issue or ideology depending on the uncertainty of individuals in accepting opinions not closer to theirs. On applying the model to simulate radicalization,  a proportion of the online population, subject to extremist propaganda radicalize depending on their pre-conceived opinions and opinion uncertainty. It is found that an optimal counter propaganda that prevents radicalization is not necessarily centrist.
\end{abstract}

\section{Introduction}
\begin{figure}[h!]
\begin{centering}
\includegraphics[width=0.5\textwidth]{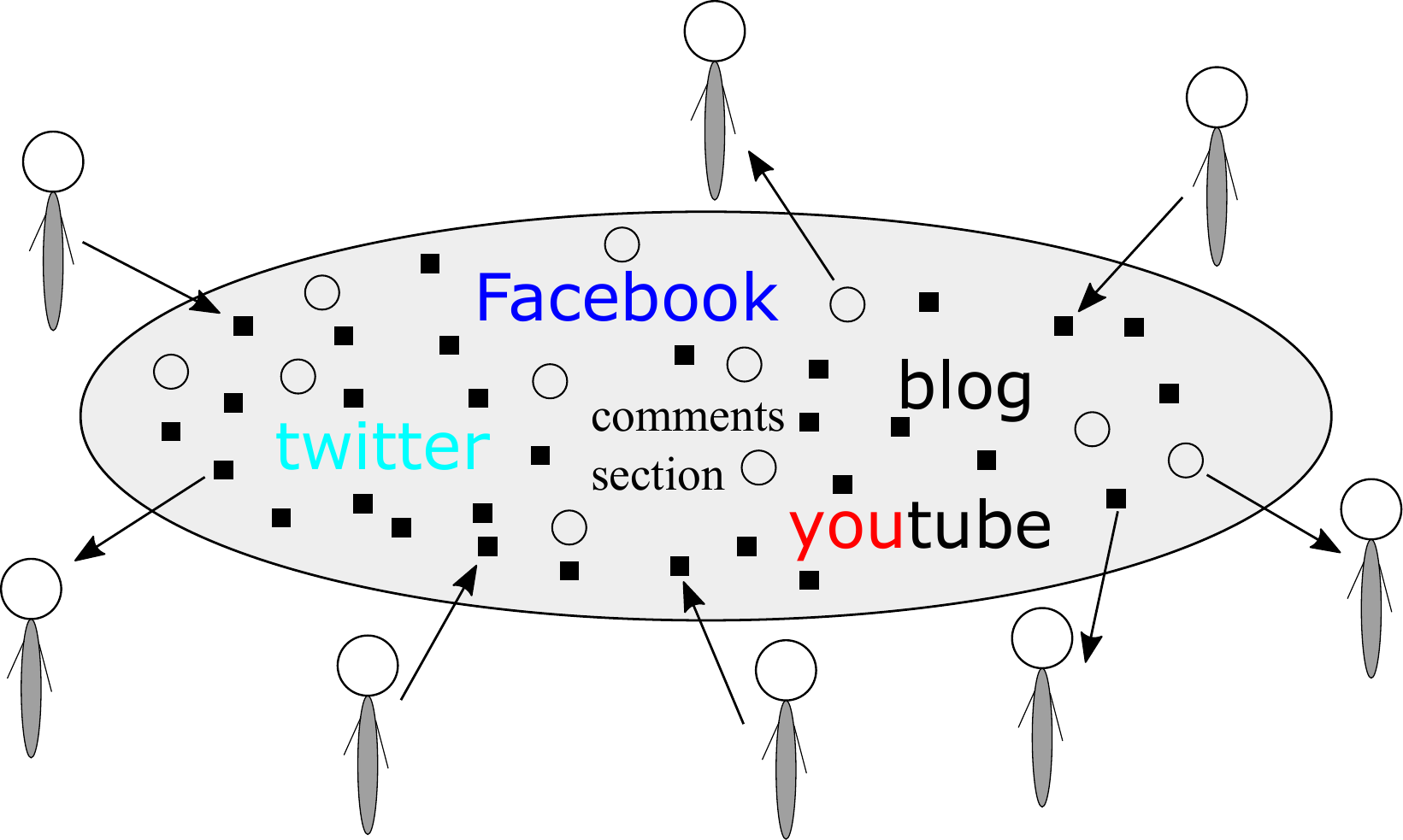}
\par\end{centering}
\caption{Individuals interacting (and influencing each other) through the internet by accessing and creating content (black squares). Shown also is individuals accessing propaganda (circles).  \label{fig:inter}}
\end{figure}

Individuals not only consume information from mass media such as television, radio, newspapers etc. but also create information and communicate through the world wide web (see illustration Fig.\ref{fig:inter} ). Information diffusing through this system strongly influences the opinion formation mechanism of an individual depending on how uncertain the individual is, regarding  his or her current opinion on a certain issue. Depending on this uncertainty, individuals tend to accept or avoid opinions that confirm or contradict  their respective pre-conceived opinions. This is a characteristic of a major cognitive bias, the {\emph{confirmation bias}} \cite{nickerson1998confirmation}. Moreover, an individual or groups of individuals can be targeted with a certain propaganda to specifically influence their opinions to satisfy a certain monetary or an ideological objective. While propaganda aimed at an individual or groups of individuals to sell a certain product for monetary gain is an acceptable common marketing strategy \cite{cialdini2001harnessing}, ideological propaganda such as that of ISIS (e.g. \cite{Berger2015}) aimed at radicalization to violent extremism poses a grave threat to the safety and security of the population \cite{Borum2011a}. While it is impossible to completely eliminate such extremist propaganda, there has been intense activity on developing counter-narratives to counter such extremist propaganda \cite{Thomas2009,Nasser-Eddine2011,Bjelopera2012,Watts2015,Ferguson2016}.

In this paper,  the opinion dynamics of a certain population of  individuals active online and are subject to a certain extremist propaganda is investigated and the effects of various counter-narratives (opinions) analyzed. The model proposed in this paper is based on the 'Bounded Confidence' method (BCM) \cite{deffuant2000mixing,Hegselmann2002,Deffuant2002}, that  simulates opinion dynamics in populations within the framework of 'sociophysics' \cite{galam2004sociophysics} (See \cite{castellano2009} for an overview on various paradigms for modeling opinion dynamics).

Of interest in these models is the overall behavior of a certain population of several individuals (or agents) that interact with each other according to certain pre-defined rules of interaction specified at the level of the individual. Characteristic of BCM is  a) a mapping of the individuals' opinions on a continuous scale between extremes and b) they influence each other if and only if the 'distance' between their opinions is smaller than a certain threshold value of uncertainty (characterized by a parameter that bounds the confidence level on their current opinion), See \cite{lorenz2007} for a review of BCM. 
The earliest work on simulating opinion dynamics within the framework of BCM goes back to the work of \cite{Hegselmann2002,Deffuant2002,deffuant2006}. They investigated the role of extremists and the variation of the uncertainties of the population on the overall opinion dynamics. The dynamic was governed by pair-wise interactions of individuals who modify their opinions depending on a certain function of the individual uncertainties. The function that relates the uncertainties and the opinions can also be designed to reflect the emotions associated with the opinions \cite{Sobkowicz2015}. However, assuming fixed uncertainties,  \cite{Mathias2016} showed that emergence of fluctuating opinions of moderates for large uncertainties. Using BCM and assuming that a certain proportion of individuals are anti-conformist, \cite{weisbuch2015}  incorporate the anti-conformist model of \cite{smaldino2015social} into the BCM to investigate appropriate conditions conducive for extremism even if the initial opinions of all the individuals are moderate. Similar in scope is the model proposed by \cite{Fan2015} based on social judgment theory, wherein in addition to compromise of similar opinions that lie within the uncertainty threshold, repulsion of opinions that are highly dissimilar is considered. These interactions of opinions on a continuous space is combined with discrete choices to take into consideration how individuals perceive the opinions of others. With regards to propaganda, \cite{Carletti:06} investigated conditions under which propaganda modifies the opinions using BCM.   

Inspired by the aforementioned models and subscribing to the philosophy of keeping things as simple as possible, the proposed model assumes that the uncertainties of all individuals are uniform and constant. The interaction of the individuals is not pair-wise but random. In essence, this is more suited to the situation where individuals online, access information that reflects the opinion of others  instead of actual inter-personal contact. The opinion update is simultaneous and done in parallel. The model is elaborated in detail in the next section. The rest of the paper is structured as follows. Following the model description, the results of three computational experiments on opinion formation with and without propaganda are presented. Finally, a discussion and interpretation of the results and conclusions provided.
\section{Model}
Let $\mathsf{O=\bigcup _{i=1}^{N}o_i}$ be the set of all opinions in a certain virtual population consisting of $N$ individuals. Irrespective of the issue under debate, the individual opinions are assumed to be distributed continuously between two extremes. In our case,  these extreme opinions are assumed to be $\mathsf{o_{E-}=0}$ and $\mathsf{o_{E+}=1}$ such that all opinions in $\mathsf{O}$ are equal to or lie between these limits. The opinions $\mathsf{o_i=0.5}$ are assumed to be a moderate (centrist) opinion.  Assuming the initial opinions in $\mathsf{O}$ are randomly distributed, the temporal evolution $\mathsf{O(t)}$ of the opinions are influenced by interactions among the individual opinions in the population and/or subject to a certain set $\mathsf{P=\bigcup _{k=1}^{N_p}o_{p_k}}$ of $\mathsf{N_p}$ external propaganda. It is also assumed that the individuals in the population do not influence the set of propaganda $\mathsf{P}$.

The mechanism of influence of individual opinions on each other is based on a certain available heuristic, the confirmation bias in human decision making. The individuals in the population are only influenced by opinions that are 'close-enough' to their pre-existing opinion characterized by a certain threshold $\mathsf{u}$, i.e. if an individual at time $\mathsf{t}$ with a certain opinion $\mathsf{o_i(t)}$ accesses the opinion $\mathsf{o_j(t)}$ of another individual in the population through online interactions i.e reading a Facebook post, twitter post or a comment on one of the multitude of discussion or comment sections online (see Fig.\ref{fig:inter}), then the individual modifies his opinion and at discrete future time becomes $\mathsf{o_i (t+1)= o_i + w_{ij}(o_j(t)-o_i(t))}$, if and only if the magnitude of the difference in opinions $\mathsf{d_{ij}=\left|o_i-o_j\right|<u}$. If this is not the case, the opinion of the individual is not affected, $\mathsf{o_i(t+1)=o_i(t)}$. The manner in which the individuals access each others' opinion is assumed to be random. 

\begin{figure}[htb]
\begin{centering}
\includegraphics[width=\textwidth]{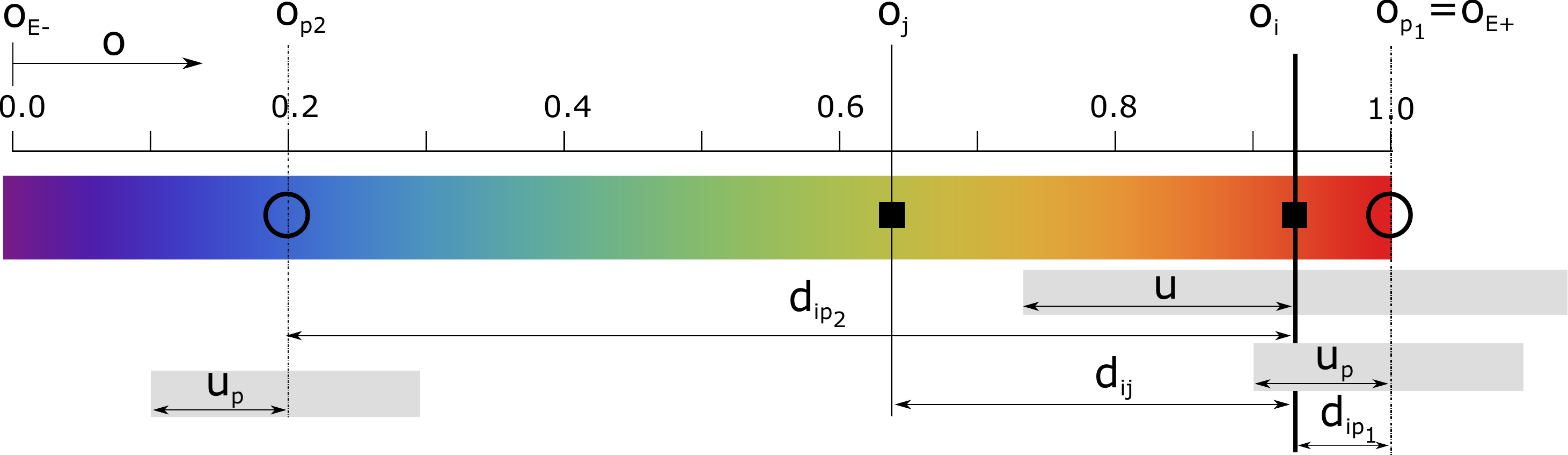}
\par\end{centering}
\caption{The opinion space $\mathsf{0\leq o \leq 1}$ and the corresponding color code that is used in this paper. Illustrated is all the relevant parameters that influence an opinion  $\mathsf{o_i}$ who has randomly accessed an opinion $\mathsf{o_j}$ in the presence of two propaganda, an extremist propaganda $\mathsf{o_{p_1}=1}$ and a propaganda $\mathsf{o_{p_2}=0.2}$ assuming $\mathsf{u_p=0.1}$  and $\mathsf{u=0.2}$. Here,  $\mathsf{o_i}$ is only influenced by $\mathsf{o_{p_1}}$.   \label{fig:leg}}
\end{figure}

To model the the influence of a certain propaganda $\mathsf{o_{p_k}\in P}$ on the individuals' opinions in the population, the opinion $\mathsf{o_i(t)}$ is now modified at time $\mathsf{t+1}$ as $\mathsf{o_i(t+1)=o_i(t)+w_{i_{p_k}}(o_{p_k}-o_i(t))}$ if and only if  $\mathsf{d_{ip_k}=\left|o_i-o_{p_k}\right|<u_p}$. Here $\mathsf{u_{p}}$ is the threshold value below which the propaganda can influence a certain individual whose opinion distance to the propaganda is $\mathsf{d_{ip_k}}$. Fig.\ref{fig:leg} illustrates all the parameters involved in the model.  

The threshold $\mathsf{u}$ can be interpreted as a certain characteristic of the whole population with regards to how flexible the individual is in modifying his pre-existing opinion or how uncertain the individual is, about his or her current opinion. This can range from $\mathsf{u=0}$ for individuals who would not change their pre-existing opinion no-matter what, to $\mathsf{u=5}$ for individuals who accept opinions that are extreme to their current opinion. The weights are chosen as $\mathsf{w_{ij}=w_{ip_k}}$ i.e. the individual at each discrete time increment averages his opinion with that of the content he is accessing online (user-generated) and all external propaganda. The weights must satisfy $\mathsf{w_{ij} +\sum_k w_{i_{pk}}=1}$. Let  $\mathsf{w_{ij}=w_{ip_k}=\frac{1}{2+N_p}}$ with $\mathsf{N_p}$ denoting the number of propaganda.  At each discrete time step, if a certain opinion is within the uncertainty of the individuals' current opinion, then the individual averages his opinion with that of the accessed opinion.

Having adequately developed all the components of the model, the generalized equation for opinion update that is applied in parallel to all opinions in $\mathsf{O}$ consisting of $\mathsf{N}$ individuals subject to $\mathsf{k}$ propaganda taking into account mutual interactions characterized by the thresholds $\mathsf{u}$ and $\mathsf{u_{p_k}}$ is given by: 
\begin{equation}
\mathsf{o_{i}(t+1)=o_{i}(t)+w_{ij}(o_{j}(t)-o_{i}(t))H\left[u-d_{ij}\right]+\sum_{k=1}^{N_{p}}w_{ip_{k}}(o_{p_{k}}-o_{i}(t))H\left[u_{p}-d_{ip_{k}}\right]\label{eq:gen}}
\end{equation}
The {\sc Heaviside} step function, $\mathsf{H}$ in Eq.(\ref{eq:gen}) is defined as follows:
\begin{equation}
\mathsf{H[n]=\left\{ \begin{array}{cc}
0 & n\leq0\\
1 & n>0
\end{array}\right.}\label{eq:heaviside}
\end{equation}
\subsection{The population-parameter space}
All the results of the computational experiments in this paper will be illustrated using the population parameter space. For a certain parameter that could be the state of the opinions after a certain number of discrete time steps $\mathsf{t}$ or a certain uniform threshold value $\mathsf{u}$ or a certain propaganda $\mathsf{u_{p_k}}$, the set of all opinions $\mathsf{O}$ is first sorted from 0 to 1 in increasing values and color-coded according to the legend shown in Fig.\ref{fig:leg}.   
\section{Computational Experiments}
In all the computational experiments presented in this paper the population $\mathsf{N=25000}$ individuals and are subject to a maximum of 2 propaganda, $\mathsf{o_{p_1}}$ and $\mathsf{o_{p_2}}$. A fixed extremist propaganda with $\mathsf{o_{p_1}=1}$ and a variable propaganda $\mathsf{0 \leq o_{p_2}\leq 1}$ with $\mathsf{u_p=1}$. 
\subsection{Opinions of a population as a result of online interactions}
\begin{figure}[h!]
\begin{centering}
\includegraphics[width=\textwidth]{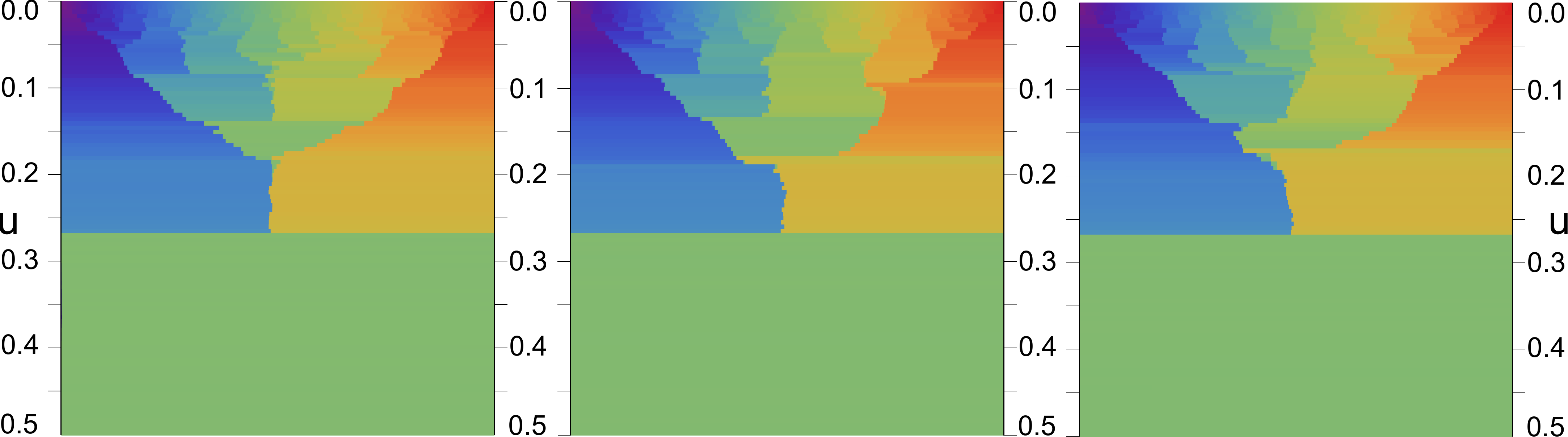}
\par\end{centering}
\caption{Map of the proportion of opinions in a population. Shown is the influence of the uncertainty parameter $\mathsf{u}$ on the steady state opinions of $\mathsf{25000}$ individuals with an initial random distribution of opinions (For color code refer to Fig.\ref{fig:leg}).  \label{fig:noext}}
\end{figure}

The first simulation is regarding the long-term opinion formation of a population with 25000 individuals  whose initial opinion is assumed to be randomly distributed. The population is not subject to any external propaganda. Fig.\ref{fig:noext} shows maps of 3 random realizations of the long-term opinions (convergent opinions after a certain time) of the population as a function of the opinion uncertainty. Depending on the uncertainty, seen is the development and formation of homogeneous sub-clusters with a uniform opinion i.e. formation of groups of individuals who subscribe to a uniform opinion. The number of clusters is inversely proportional to the  uncertainty $\mathsf{u}$. There is no novelty in these predictions as these results are already well known albeit for a slightly different pair-wise interaction. 
For $\mathsf{u> \approx 0.26}$, there is a single cluster with uniform consensus. Not visually discernible is a very small negligible proportion of extremists for $\mathsf{u>0.26}$. If the population is uncertain on a certain issue, without any propaganda, assuming idealized interactions described in the previous section, there is consensus of opinion and this opinion is moderate. The majority of the population is in harmony except for a very small proportion of fringe elements that hold extreme opinions. For smaller uncertainties, the population is polarized on the same issue keeping all other factors constant. For smaller uncertainties there is a fragmentation of opinions into 3, 4, 8 ... etc. sub-clusters. The type of opinion is denoted by the color whose details are provided in Fig.\ref{fig:leg}. 
\subsection{Opinions of a population as a result of online interactions with extremist propaganda}
\begin{figure}[!h]
\begin{centering}
\includegraphics[width=\textwidth]{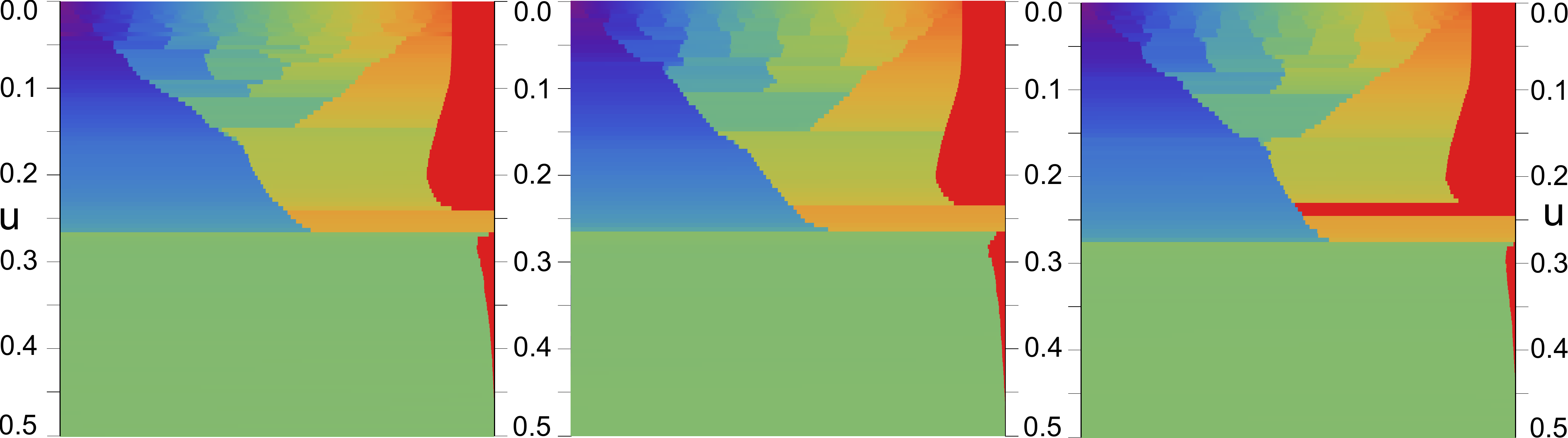}
\par\end{centering}
\caption{Map of the proportion of opinions in a population with extremist propaganda (case 2). Shown is the influence of the uncertainty parameter on the long term opinions of 25000 individuals with an initial random distribution of opinions showing unstable behavior around the population uncertainty range $\mathsf{0.23 < u<0.24}$ (For color code refer to Fig.\ref{fig:leg})}
\label{fig:ext1}
\end{figure}

In the second simulation, the influence of one extremist propaganda on the long-term opinion formation in the population is taken into consideration. As in the first simulation, the strength of the population is 25000 and the initial opinion of the population is assumed to be randomly distributed. The propaganda is assumed to subscribe to the opinion $\mathsf{o_{p_1}=1}$ i.e. an extremist opinion (e.g. hate preachers, ISIS propaganda etc...).  Two special cases are considered:
\begin{enumerate}
\item In case 1, the extremist propaganda targets individuals with opinions $\mathsf{o_i \geq 0.95}$
\item In case 2, the extremist propaganda targets individuals with opinions  $\mathsf{o_i\geq 0.9}$
\end{enumerate}
In practical terms, case 2 propaganda can be interpreted as being more efficient than case 1, i.e. an individual with an opinion $\mathsf{o=0.92}$ will not be influenced by the propaganda according to case 1 but will be influenced by the propaganda according to case 2.
\begin{figure}[h!]
\begin{centering}
\includegraphics[width=\textwidth]{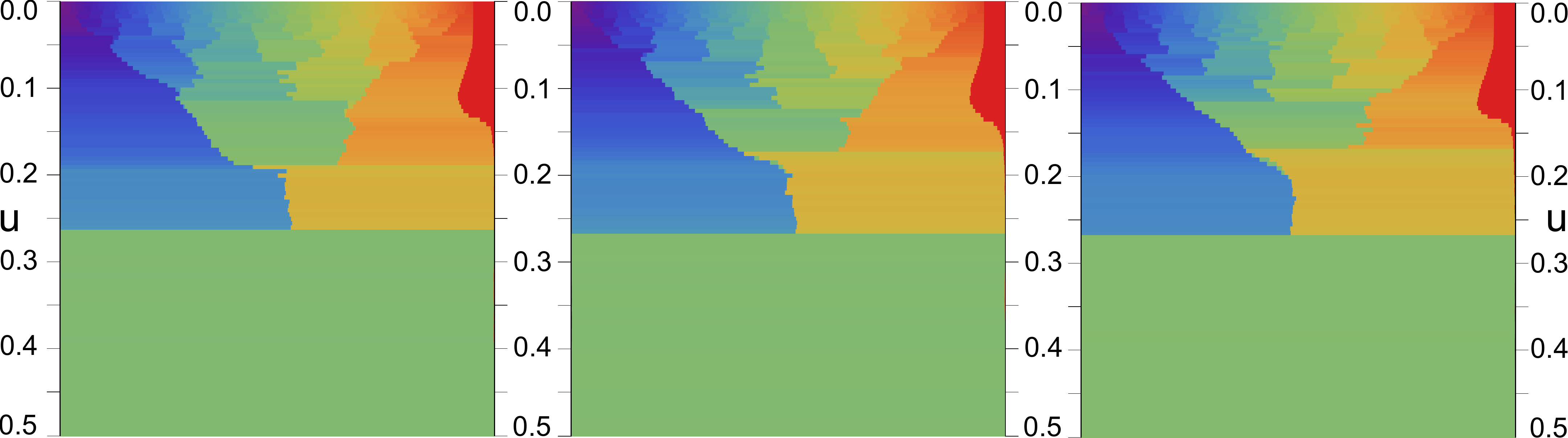}
\par\end{centering}
\caption{Map of the proportion of opinions in a population with extremist propaganda (case 1) Shown is the influence of the uncertainty parameter on the long term opinions of 25000 individuals with an initial random distribution of opinions  (For color code refer to Fig.\ref{fig:leg}). 
\label{fig:ext05}}
\end{figure}
3 random realizations of the map of the long term opinions according to case 1 is shown in Fig.\ref{fig:ext05}. In contrast to the predictions in Fig.\ref{fig:noext}, there is a certain range of uncertainties of the population $\mathsf{0\leq u< 0.15}$ wherein there is a formation of a cluster of individuals with a uniform extremist opinion. This is to be expected in the presence of extremist propaganda. According to case 2, (See Fig.\ref{fig:ext1})  there is an even larger range of uncertainties that allow for the formation of a uniform cluster of extremist opinions consistent with that of the propaganda. The overall proportion of extremists with opinions consistent with that of the propaganda are higher for case 2 than for case 1.  In case 2, as seen in Fig,\ref{fig:ext1}, the proportion of extremists is constant with a value of 10\% of the total population until the uncertainty of the population is 0.1. As the uncertainty of the population increases, the proportion of extremists doubles to about 20\% and at around u=0.25$\mathsf{\pm 0.2}$ there is instability  that shows either complete drop in the proportion of extremists to 0\% (see Fig.\ref{fig:ext1} left and middle) or a dramatic rise in the proportion of extremists to about 50\% of the population (see Fig.\ref{fig:ext1} right).  For a population uncertainty of about u=0.25$\mathsf{\pm 0.2}$, the proportion of extremists is unpredictable. Anything is possible. Either 50\% of the population become extremist or extremism is completely rooted out of the population. To provide a definitive origin for this instability is beyond the scope of this paper (see also \cite{hegselmann2015}).   
For uncertainties greater that $\mathsf{u=0.275}$, there is overall consensus with a small proportion of extremists.
\subsection{Opinions of a population as a result of online interactions with, an extremist propaganda and a counter-propaganda}
In the final simulation, in addition to an extremist propaganda with $\mathsf{o_{p_1}=1}$, the long term opinions of individuals in the population taking into account online interactions and a counter-propaganda  $\mathsf{0 \leq \mathsf{o_{p_2}\leq 1}}$ is considered. Of interest is the effect of this counter propaganda and its opinion $\mathsf{o_{p_2}}$ on the proportion of extremists in the population as a function of the population uncertainty.

\begin{figure}[htb]
\begin{centering}
\includegraphics[width=\textwidth]{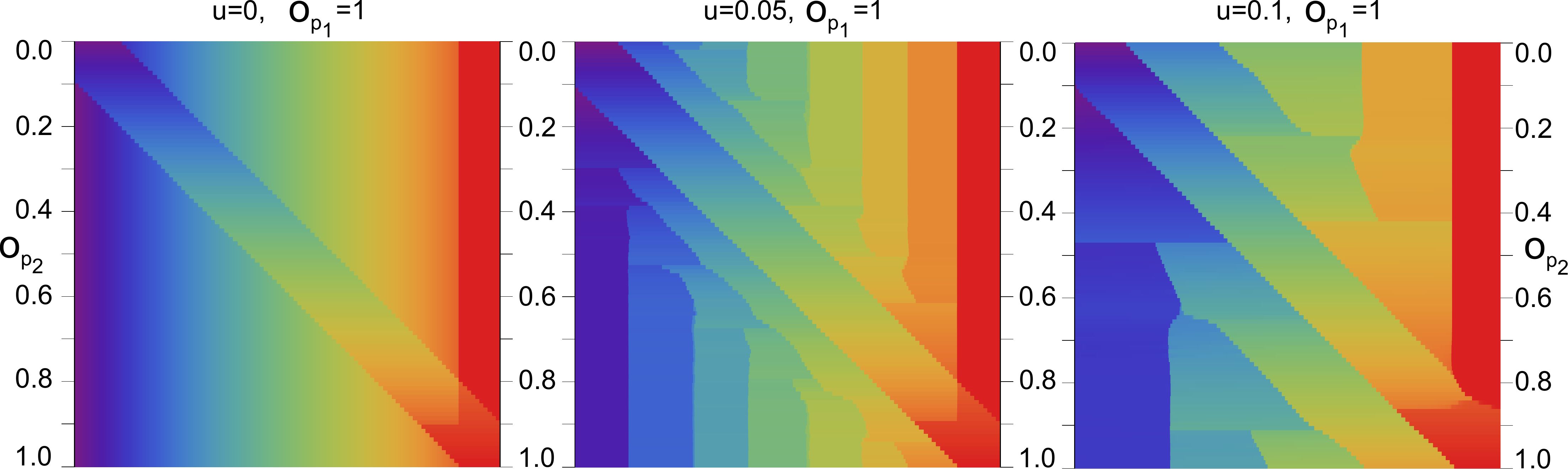}
\par\end{centering}
\caption{Map of the influence of the target propaganda $\mathsf{o_{p_2}}$ in the presence of propaganda $\mathsf{o_{p_1}=1}$ on the steady state opinions of $\mathsf{25000}$ individuals with an initial uniform distribution of opinions with extremist propaganda with $\mathsf{u_p=0.1}$ for three different uncertainties of the population. Left: $\mathsf{u=0}$, middle:$\mathsf{u=0.05}$ and right: $\mathsf{u=0.1}$ (For color code refer to Fig.\ref{fig:leg}) \label{fig:op1}}
\end{figure}

\begin{figure}[htb]
\begin{centering}
\includegraphics[width=\textwidth]{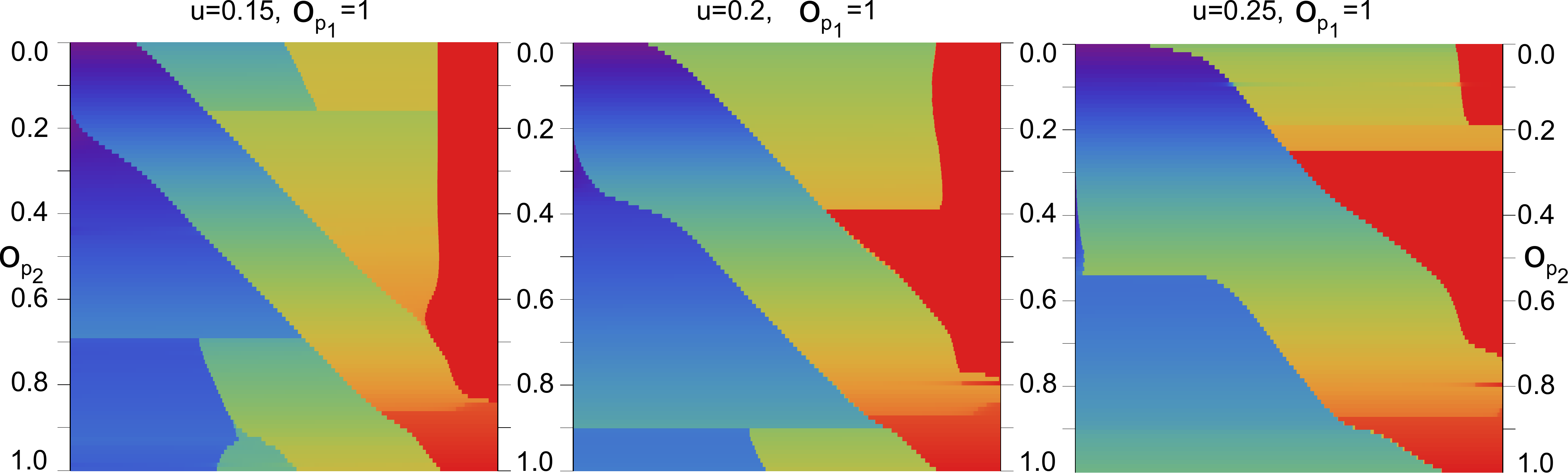}
\par\end{centering}
\caption{Map of the influence of the target propaganda $\mathsf{o_{p_2}}$ in the presence of propaganda $\mathsf{o_{p_1}=1}$ on the steady state opinions of $25000$ individuals with an initial uniform distribution of opinions with extremist propaganda with $\mathsf{u_p=0.1}$ for three different uncertainties of the population. Left: $\mathsf{u=0.15}$, middle:$\mathsf{u=0.2}$ and right: $\mathsf{u=0.25}$  (For color code refer to Fig.\ref{fig:leg})\label{fig:op2}}
\end{figure}

\begin{figure}[htb]
\begin{centering}
\includegraphics[width=\textwidth]{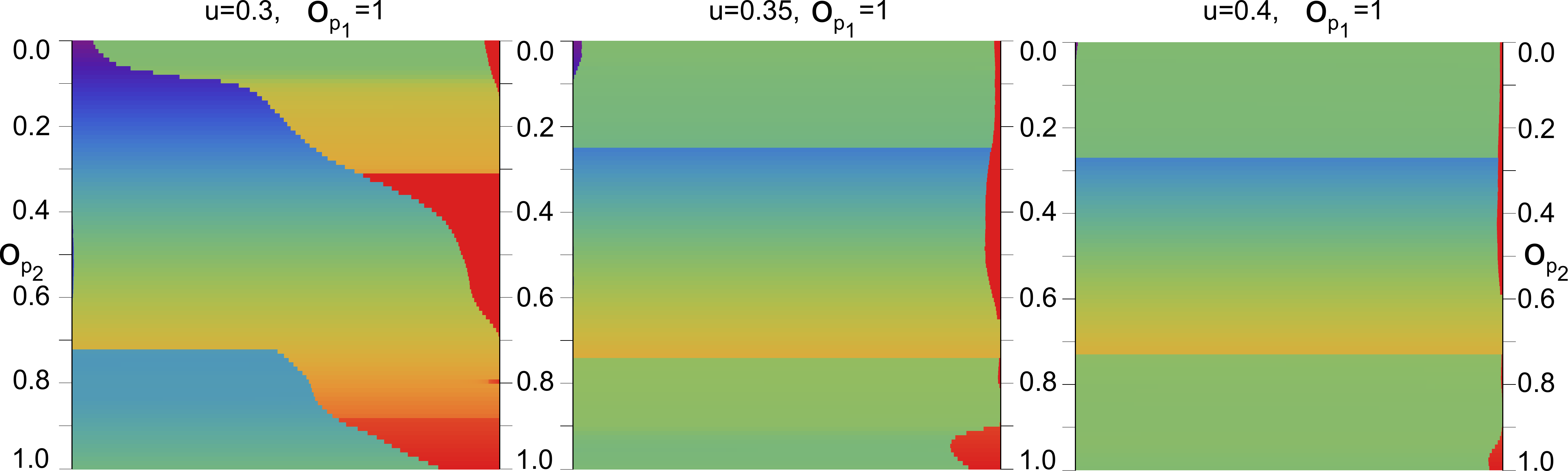}
\par\end{centering}
\caption{Map of the influence of the target propaganda $\mathsf{o_{p_2}}$ in the presence of propaganda $\mathsf{o_{p_1}=1}$ on the steady state opinions of $25000$ individuals with an initial uniform distribution of opinions with extremist propaganda with $\mathsf{u_p=0.1}$ for three different uncertainties of the population. Left: $\mathsf{u=0.3}$, middle:$\mathsf{u=0.35}$ and right: $\mathsf{u=0.4}$  (For color code refer to Fig.\ref{fig:leg})\label{fig:op3}}
\end{figure}

Figs.\ref{fig:op1}, \ref{fig:op2} and \ref{fig:op3} show the map of the long-term opinions for various uncertainties $\mathsf{u}$ and various counter-propaganda $\mathsf{o_{p_2}}$. For $\mathsf{u=0}$, according to Fig.\ref{fig:op1} (left), shows a constant proportion ($\mathsf{10\%}$) of extremists irrespective of the target propaganda $\mathsf{o_p}$ for $\mathsf{o_p<0.9}$. For $\mathsf{o_p>0.9}$, there is a a sudden increase in the proportion of extremists to $\mathsf{20\%}$ of the total population that linearly decreases to $\mathsf{10\%}$ of the total population. The opinion space is highly fragmented as expected from the assumption $\mathsf{u=0}$. A similar trend regarding the proportion of extremists is observed in Fig.\ref{fig:op1}(middle) for $\mathsf{u=0.1}$ except that the overall population is fragmented into clusters in terms of opinions. For larger uncertainty, $\mathsf{u=0.15}$, there is a) less fragmentation of opinions, b) absence of extremist opinion $\mathsf{o_i=0}$ for $\mathsf{o_p>0.3}$ and c) a marginal increase of extremist opinions at around $\mathsf{25\%}$ for $\mathsf{o_p=0.9}$ that linearly decreases to $\mathsf{\approx 10 \%}$. However, just before  $\mathsf{o_p=0.9}$ there is almost $\mathsf{50\%}$ decrease in the number of extremists before a drastic rise in the extremist proportion. This effect is observed in a much more amplified manner for $\mathsf{u=0.15}$, $\mathsf{u=0.2}$ and $\mathsf{u=0.25}$ depicted in Fig.\ref{fig:op2},(left, middle and right) respectively. The opinions in general are a mix of fragmentation and polarization with extremists. 

Of significance is the absence of extremists at around $\mathsf{o_{p_2}\approx 0.85}$ for $\mathsf{u=0.25}$, $\mathsf{u=0.2}$ and also for $\mathsf{u=0.15}$ but only for a much smaller range of  $\mathsf{o_{p_2}}$. Such an absence of extremists is also found for $\mathsf{o_{p_2}\approx 0.25}$ for  $\mathsf{u=0.25}$ and $\mathsf{u=0.3}$ as seen in Figs.\ref{fig:op2}(right) and Figs.\ref{fig:op3}(left). For large values of population uncertainties, $\mathsf{u=3.5}$, and $\mathsf{u=0.4}$, there is predominant consensus for the complete range of  $\mathsf{o_{p_2}}$ with the consensus of the majority swaying between moderate opinions. 


\begin{figure}[htb]
\begin{centering}
\includegraphics[width=\textwidth]{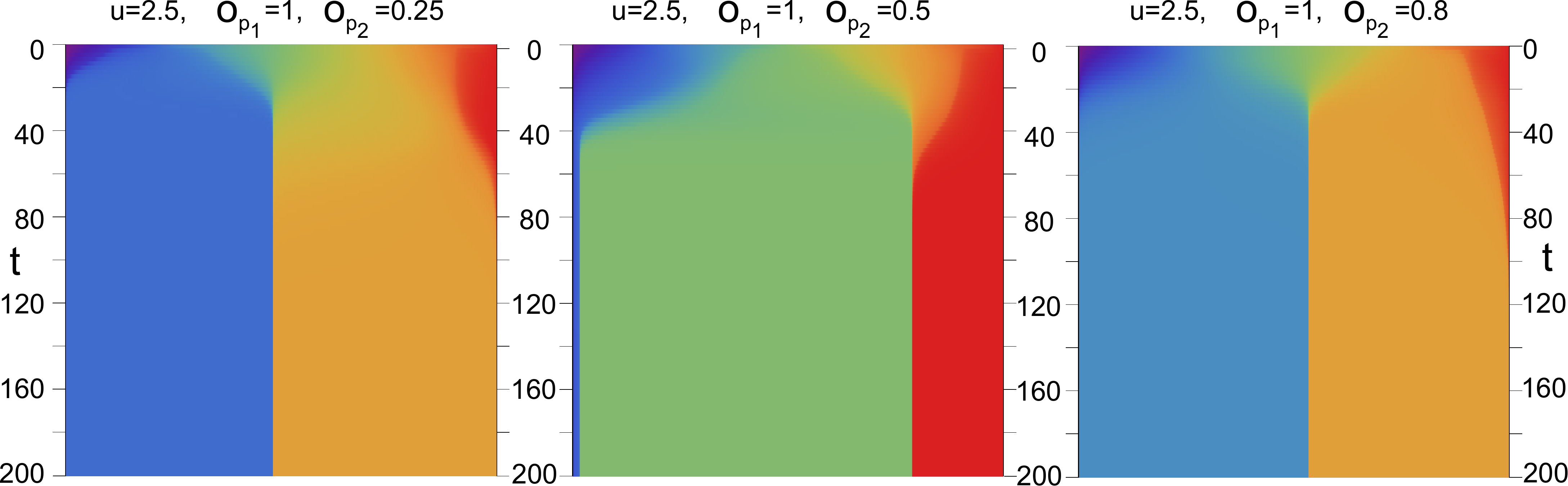}
\par\end{centering}
\caption{Map of the temporal evolution of opinions for $\mathsf{o_{p_2}=0.29}$ (left), $\mathsf{o_{p_2}=0.5}$ (middle) and $\mathsf{o_{p_2}=0.8}$ (right) in the presence of extremist propaganda $\mathsf{o_{p_1}=1}$ for 25000 individuals with an initial random distribution of opinions assuming the population uncertainty $\mathsf{u=0.25}$  (For color code refer to Fig.\ref{fig:leg})\label{fig:temporal}}
\end{figure}

All the results so far have been long-term opinions that have converged to a constant opinion after several iterations of Eq.\ref{eq:gen}. Fig.\ref{fig:temporal} shows the temporal evolution of opinions with an initial random distribution for the case of $\mathsf{u=0.25}$ for $\mathsf{o_{p_2}=0.2}$ (left), $\mathsf{o_{p_2}=0.5}$ (middle) and  $\mathsf{o_{p_2}=0.8}$ (right). For the same population uncertainty, as already seen within the context of the long-term opinion formation, the propaganda $\mathsf{o_{p_2}=0.2}$ shown in Fig.\ref{fig:temporal} (left), and  $\mathsf{o_{p_2}=0.8}$ Fig.\ref{fig:temporal} (right) eliminate extremism completely from the population even with extremist propaganda of similar strength. However, for a propaganda $\mathsf{o_{p_2}=0.5}$ depicted in Fig.\ref{fig:temporal}(middle) shows an increase in extremism in time.
\section{Discussion}
In the first computational experiment the long term-opinion formation in the population is dependent on the initial opinion and the uniform uncertainty of the population. The opinions of individuals in the population is influenced by online interactions (e.g. opinionated creation and access of content online). The major observation is the formation of clusters, i.e. groups of individuals with the same opinions. The number of clusters depends on the uncertainty of the population. High uncertainties generate consensus in the population whose opinion is that of a center/moderate opinion while low uncertainties support the formation of multiple groups ranging the complete opinion space.  Assuming the  population to have a uniform uncertainty can be argued to be an oversimplification but it can be interpreted as a common trait of the population. We can conclude that it is desirable that the population have high levels of uncertainty in opinion regarding a certain issue to avoid conflict. 

In the second computational experiment, an external extreme propaganda was introduced as an additional source of influence on the long-term opinion of the population. The extremist propaganda is targeted at the extreme individuals in the population. Depending on the population uncertainty, through online interactions, these extreme individuals constantly influenced by extreme propaganda can influence other individuals in the population. Clearly, even if the propaganda is aimed at a certain section of the population, the uncertainty of the population and the presence of online interactions can influence the rest of the population. While the extreme propaganda is the source, the online interaction acts as a tool to spread this extreme propaganda. The strength of the propaganda was also investigated, showing a proportional influence on the long term extremist proportions in the population. The most interesting result according to Fig.\ref{fig:ext1} is that of the complete absence of extremists in the population for a certain narrow range of population uncertainty. 

The final experiment can be interpreted as a scenario wherein a certain agency wants to find out which is the optimal counter strategy to eliminate or curb extremist opinions in a certain population in constant online interactions and are subject to a certain extremist propaganda. Such counter measures are crucial to curb violent extremism committed by certain individuals holding radical extremist opinions. Within this context, the simulations evidently show that there is indeed certain ranges of opinions that in terms of propaganda can actually counter extremist propaganda by reducing the number of individuals that subscribe to extremist opinions. These counter propaganda, according to computational simulations show that it is indeed not a moderate opinion but an opinions that lies midway in-between extreme and moderate opinions depending on the uniform uncertainty of the population.  

\section{Conclusion}

The opinion dynamics of an ideal population subject to propaganda is simulated using numerical computations.  The computational model is based on the bounded confidence approach. 
In the absence of propaganda, individuals with diverse initial opinions modify their opinions due to mutual interactions. The change in opinion is based on their uncertainty (i.e the distance of an opinion from his/her current opinion that the individual is ready to compromise with) with regard to the issue. Depending on this uncertainty, the population  is fragmented and in conflict for low uncertainties or develop consensus for large uncertainties. In the presence of an external extremist propaganda aimed at the extreme elements in this population, mutual interactions within the population can transform non-extremist individuals into ones that hold extremist opinions.  To counter this extremist propaganda, a counter-propaganda with various opinions ('narratives') was introduced showing that the most optimal counter-strategy to curb individual opinion shift to extremism is neither the complete opposite nor a centrist propaganda.

\bibliographystyle{apalike}
\bibliography{OpD2} 

\end{document}